\documentclass[%
 reprint,
superscriptaddress,
 amsmath,amssymb,
 aps,
]{revtex4-2}

\usepackage{graphicx}
\usepackage{dcolumn}
\usepackage{bm}
\usepackage{hyperref}
\usepackage{color}

\begin{document}

\preprint{APS/123-QED}

\title{Deterministic preparation of squeezed and non-Gaussian states in coupled qubit-boson systems}

\author{Fengquan Zhang}%
\thanks{These authors contributed equally to this work.}
\affiliation{State Key Laboratory of Precision Spectroscopy, School of Physics, East China Normal University, Shanghai 200241, China.}
\author{Xiaotong Liu}%
\thanks{These authors contributed equally to this work.}
\affiliation{State Key Laboratory of Precision Spectroscopy, School of Physics, East China Normal University, Shanghai 200241, China.}
\author{Yuping Mao}
\affiliation{School of Science, Huzhou Normal University, Huzhou, Zhejiang 313000, China}
\author{Zhiyuan Lin }
\affiliation{State Key Laboratory of Precision Spectroscopy, School of Physics, East China Normal University, Shanghai 200241, China.}
\author{Tim Byrnes}%
 \email{tim.byrnes@nyu.edu}
\affiliation{New York University Shanghai; NYU-ECNU Institute of Physics at NYU Shanghai,
567 West Yangsi Road, Shanghai, 200124, China.}
\affiliation{State Key Laboratory of Precision Spectroscopy, School of Physics, East China Normal University, Shanghai 200241, China.}
\affiliation{Center for Quantum and Topological Systems (CQTS),
NYUAD Research Institute, New York University Abu Dhabi, UAE.
 }
\affiliation{Department of Physics, New York University, New York, NY 10003, USA.}
\date{\today}

\begin{abstract}
Deterministic preparation of bosonic resource states is a central task in
continuous-variable quantum information processing and quantum metrology. Here we develop a measurement-based imaginary time evolution (MITE) protocol for preparing
squeezed and non-Gaussian states. The protocol combines
ancilla-mediated weak measurements of the Hamiltonian with a conditional phase rotation that suppresses excited state trapping
and stabilizes convergence toward the ground state. The method is potentially applicable in hybrid discrete variable--continuous variable (DV-CV) platforms, where the qubit acts as the measurement readout.  We demonstrate
deterministic convergence over a broad range of parameters and
show that the performance degradation near the stability boundary is
governed by the shrinking Bogoliubov quasiparticle gap. The same framework
also extends to non-Gaussian bosonic resource states, including
Schr\"odinger cat states and excited states of the single-mode squeezing Hamiltonian, which display Wigner negativity.  The squeezed states remain robust
under bosonic loss over experimentally relevant regimes.  
\end{abstract}

\maketitle


\section{\label{sec:level1}INTRODUCTION}

Accessing ground states of quantum Hamiltonians is essential for
understanding and controlling quantum many-body systems and plays a key
role in quantum information processing, quantum simulation, quantum
computation, and quantum metrology
\cite{Feynman1982,Bloch2012,Albash2018,Giovannetti2011,Acin2018}.
This task is particularly relevant in continuous-variable bosonic
systems, where many useful resource states can be described as ground
states of suitable parent Hamiltonians
\cite{Braunstein2005,Weedbrook2012,Grimsmo2021}. Scalable and
experimentally feasible preparation of such bosonic ground states
therefore remains a central objective in bosonic quantum-state
engineering \cite{Kienzler2015,Leghtas2015,CampagneIbarcq2020}. Techniques to prepare bosonic ground states include engineered dissipation \cite{Kienzler2015}, adiabatic evolution     \cite{VenegasGomez2020}, quantum optimal control \cite{vanFrank2016}, and measurement-based protocols \cite{Lee2023}.  These approaches involve performing quantum state preparation using tailored dissipation channels, slow parameter ramps,  optimal-control sequences, or postselection, respectively. 
However, preparing a completely arbitrary bosonic state remains rather challenging due to limitations in the available controls that are available with these methods.  

Imaginary time evolution based approaches for quantum state preparation have attracted interest in recent years \cite{Magnus1954,Vidal2007,Bauer2020}.  Formally, imaginary time evolution allows for the preparation of the ground state $ |E_0 \rangle $ of a Hamiltonian $ \hat{H} $ from any state with a non-zero ground state component
\begin{align}
   \lim_{ \tau \rightarrow \infty}  e^{- \tau \hat{H} } | \psi_0 \rangle \propto |E_0 \rangle  .
   \label{imagtimeevol}
\end{align}
This process is non-unitary, and hence methods to engineer this evolution need to be devised.  Several approaches to quantum imaginary time evolution have been developed, including variational imaginary time evolution (VITE) \cite{McArdle2019,Endo2020,Jones2019,Li2024VITE_MPS}, unitary approximation methods such as quantum imaginary time evolution (QITE) \cite{Motta2020,Kamakari2022,Silva2023,Kolotouros2025}, and probabilistic imaginary time evolution (PITE) based on postselection techniques \cite{Nishi2023,Xie2024,Leadbeater2024}, and adiabatic imaginary time evolution \cite{Hejazi2024}.  These approaches approximate the intrinsically non-unitary imaginary time evolution using variational optimization, enlarged unitary dynamics, probabilistic filtering procedures, or adiabatic methods. In particular, measurement-based imaginary time
evolution (MITE) provides a physically natural route for realizing
energy-selective filtering through measurement and conditional
feedback operations
\cite{Mao2023,Kondappan2023,Chen2024AKLT,IloOkeke2022,Mao2026Stabilizer}. 
Once a target Hamiltonian is specified, MITE
provides a general, deterministic (i.e. no postselection) route toward ground-state
preparation. Such
measurement-based strategies progressively reshape the quantum state in
the Hamiltonian eigenbasis without requiring engineered dissipation
channels or complex optimal-control sequences.

\begin{figure*}[t]
\includegraphics[width=0.95\textwidth]{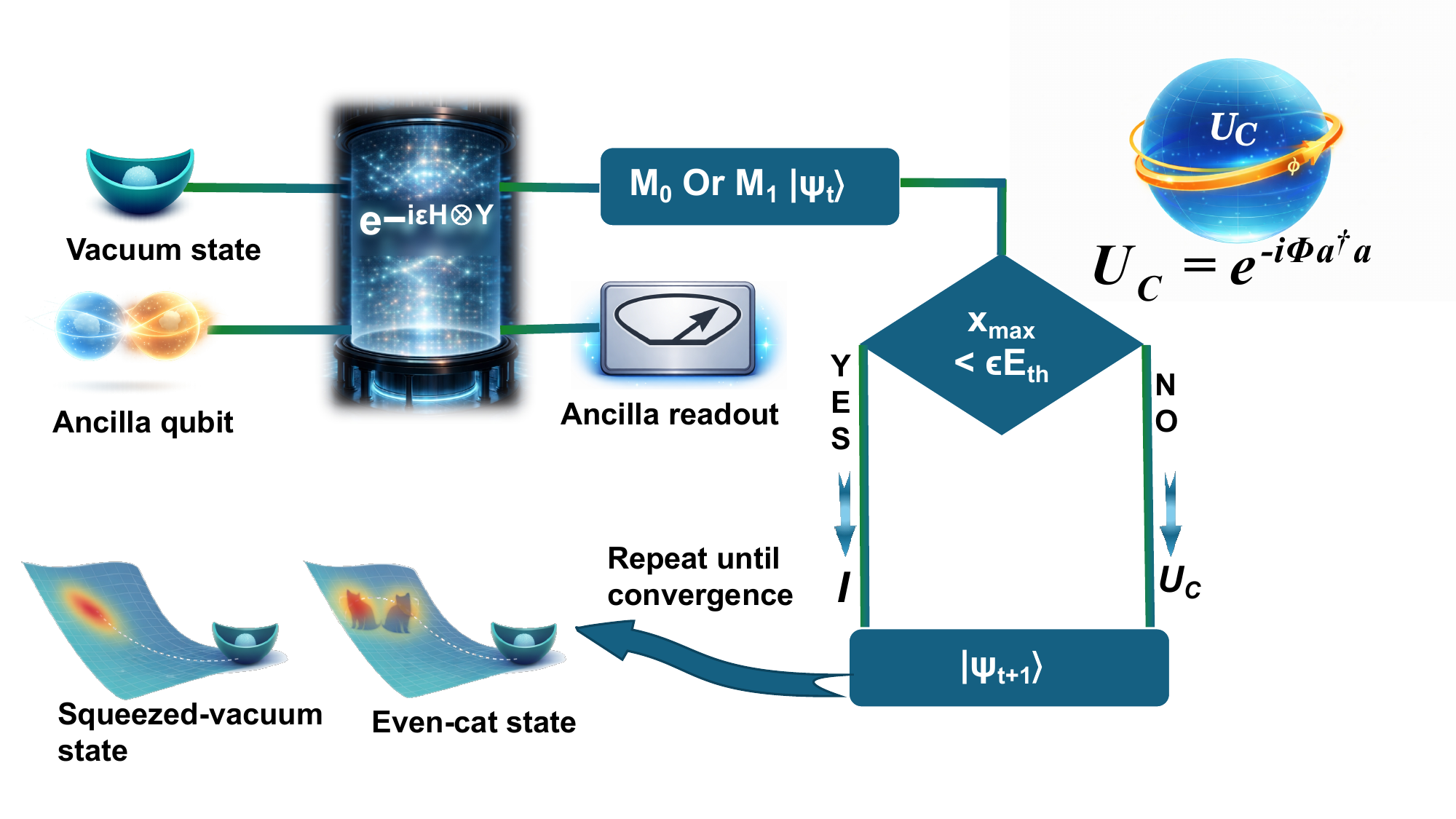}
\caption{\label{fig:fig_1}  Measurement-based imaginary time 
evolution protocol for bosonic state preparation. A bosonic mode initialized in the vacuum state is
coupled to an ancillary qubit through the interaction $U=e^{-i\epsilon \hat H\otimes Y}$. The ancilla readout induces
one of the weak measurement operators $M_0$ or $M_1$, producing the
conditional bosonic state $|\psi_t\rangle$. The accumulated measurement
record determines the filtering peak position $x_{k_0k_1}^{\max}$, which is compared with
the threshold $\epsilon E_{\mathrm{th}}$. When the threshold condition
is violated, the conditional phase rotation 
$U_C=e^{-i\phi \hat a^\dagger \hat a}$ is applied; otherwise, the weak
measurement sequence continues. Repetition of this adaptive loop
suppresses excited state trapping and drives convergence toward
the ground state.}
\end{figure*}

Bosonic degrees of freedom arise naturally in a variety of physical systems, with representative examples including electromagnetic modes at optical \cite{Glauber1963,Scully1997,Braunstein2005}  and microwave frequencies \cite{Blais2004,Schoelkopf2008}, phononic modes in mechanical resonators \cite{OConnell2010,Aspelmeyer2014}, and motional modes of trapped particles \cite{Meekhof1996,Leibfried2003}. In particular, bosonic systems have proven to be a versatile platform for performing quantum information tasks, such as quantum metrology, communication, and computation \cite{Giovannetti2011,Braunstein2005,Lloyd1999}. 
Preparing various quantum states of interest, such as squeezed states and non-Gaussian states are of fundamental interest for tasks such as quantum metrology \cite{Caves1981,Holland1993,Giovannetti2011,fadel2023multiparameter}.  In particular, negativity of the Wigner quasiprobability distribution can be viewed as a resource in continuous-variable quantum information processing \cite{Albarelli2018}. Quantum computations restricted to states, operations, and measurements with positive Wigner representations admit efficient classical simulation, implying that Wigner negativity is a necessary ingredient for quantum computational advantage \cite{Mari2012,Veitch2012,Veitch2014}.  One of the interesting developments is the creation of hybrid
discrete-variable--continuous-variable (DV--CV) architectures, where typically a qubit is coupled to a bosonic mode \cite{Jaynes1963,vanLoock2011,Andersen2015}.  
 Here, an
ancillary qubit serves primarily as the measurement interface and the
bosonic mode hosts the target continuous-variable Hilbert space. Such hybrid platforms
have already been realized in cavity QED, trapped-ion, and superconducting
circuit systems
\cite{Sayrin2011,Sun2014,Gan2020,Darras2023,Chou2024,Hou2024},
making them especially promising for implementing novel measurement-based control methods \cite{lin2026preparing}.  In the context of imaginary time evolution, previous realizations
have focused primarily on discrete-variable systems
\cite{Mao2023,Kondappan2023,Chen2024AKLT,IloOkeke2022,Mao2026Stabilizer}, while deterministic implementation in
continuous-variable bosonic platforms remains largely unexplored. A realization of this kind would establish imaginary time evolution as an effective method for 
bosonic ground-state preparation beyond purely Gaussian state-engineering
methods.

In this paper, we develop a MITE protocol for deterministic preparation
of bosonic ground states.  We show the basic idea in Fig.~\ref{fig:fig_1}. As a simple benchmarking test case, we
first consider the single-mode squeezing Hamiltonian which has a ground state that is the 
squeezed vacuum. This model illustrates how the protocol operates in a
continuous-variable system, including the roles of weak energy
measurements, adaptive feedback, spectral compression near the
stability boundary, and robustness to bosonic loss. 
Starting from an experimentally
accessible initial state, repeated weak
measurements generate an energy-selective filtering process, while
adaptive feedback prevents trapping in excited state subspaces and
stabilizes convergence toward the ground state. 
We then show that the degradation of performance near the stability
boundary is controlled by the shrinking Bogoliubov quasiparticle gap,
investigate the robustness of the prepared states against bosonic loss, and finally extend the same framework to non-Gaussian
parent Hamiltonians, including Schr\"odinger cat states. 

\section{Measurement-Based Imaginary Time Evolution}
Here we review the theory of MITE, which performs ground state preparation using measurements and feed-forward unitary operations \cite{Mao2023}.

\subsection{Construction of Ancilla-based Weak-Measurement Operators}


As given in (\ref{imagtimeevol}), by evolving in imaginary time, it is possible to prepare the ground state of any Hamiltonian $ \hat{H} $, starting from an initial state $ | \psi_0 \rangle $ with a non-zero overlap with the ground state.  However, the operator $e^{-\tau \hat H}$ is non-unitary and cannot be
implemented directly on quantum hardware.  In the case of MITE, a measurement-based approximation to imaginary time dynamics is achieved using
ancilla-based weak measurements. An ancilla qubit is initialized in
the superposition state
\begin{equation}
|+\rangle_a=\frac{|0\rangle_a+|1\rangle_a}{\sqrt{2}},
\label{eq:plusAncilla}
\end{equation}
and coupled to the bosonic mode through the joint unitary
\begin{equation}
U = e^{-i\epsilon \hat H \otimes Y},
\label{eq:jointU}
\end{equation}
where $\epsilon$ sets the measurement strength and $Y$ denotes
the Pauli-$Y$ operator acting on the ancilla.

Projective measurement of the ancilla in the computational basis
induces conditional evolution of the bosonic mode described by
\begin{equation}
M_0=\langle 0|_a\,U\,|+\rangle_a
=\frac{1}{\sqrt{2}}\left(\cos(\epsilon \hat H)-\sin(\epsilon \hat H)\right),
\label{eq:M0}
\end{equation}
\begin{equation}
M_1=\langle 1|_a\,U\,|+\rangle_a
=\frac{1}{\sqrt{2}}\left(\cos(\epsilon \hat H)+\sin(\epsilon \hat H)\right).
\label{eq:M1}
\end{equation}
In the weak-measurement regime $ || \epsilon \hat{H} || \ll 1 $, the 0 outcome approximates an
imaginary time-evolution step $M_0 \approx e^{-\epsilon \hat{H}} /\sqrt{2} $, whereas the 1 outcome produces  complementary
measurement backaction $M_1 \approx e^{\epsilon \hat{H}} /\sqrt{2} $, that biases the filtering toward higher-energy
components (see Appendix \ref{app:a} for further details).

Repeated applications of the measurement operators $M_0, M_1 $ generate a stochastic, energy-selective
filtering process in the Hamiltonian eigenbasis. In this scheme, the
ancillary qubit serves as the measurement interface, and the
weak measurements give rise to conditional
evolution of the system through projective readout of the ancilla.


\subsection{Measurement-Induced Energy Filtering under Repeated Weak Measurements}

A sequence of weak measurements acts to gradually collapse the initial state by adding a modulation envelope in the energy eigenbasis of the Hamiltonian.
Consider a particular measurement sequence consisting of $k_0$ outcomes $M_0$ 
and $k_1$ outcomes of $M_1$.  The ordering does not matter as (\ref{eq:M0}) and (\ref{eq:M1}) commute.  The action on the initial state is given by
\begin{equation}
M_0^{k_0}M_1^{k_1}|\psi_0\rangle
=
\sum_n
A_{k_0 k_1}(\epsilon E_n)
\langle E_n|\psi_0\rangle
|E_n\rangle ,
\label{k0k1state}
\end{equation}
where we defined the amplitude function
\begin{equation}
A_{k_0k_1}(x)
=
\cos^{k_0}\left(x+\frac{\pi}{4}\right)
\sin^{k_1}\left(x+\frac{\pi}{4}\right).
\label{eq:Ak0k1}
\end{equation}

The shape of the envelope is determined by $k_0, k_1 $, but for  sufficiently many measurement steps, it
approaches a Gaussian envelope in energy space (further details are given in Appendix \ref{app:a}).  
The peak position of this filtering envelope is given by
\begin{equation}
x_{k_0k_1}^{\max}
=
\epsilon E_{k_0k_1}^{\max}
=
\frac{1}{2}
\arcsin
\left(
\frac{k_1-k_0}{k_0+k_1}
\right),
\label{eq:xmax}
\end{equation}
while its width decreases with the total number of measurements as
\begin{equation}
\sigma  \simeq \frac{1}{\sqrt{2(k_0+k_1)}} .
\label{eq:sigma}
\end{equation}

Repeated weak measurements therefore sharpen the energy selectivity of
the modulation envelope and localize the state within a narrow,
measurement-record-dependent energy window.
This measurement-induced modulation envelope is the operational mechanism
underlying MITE. When the envelope is centered near the low-energy region,
higher-energy components are relatively suppressed, while low-energy
components acquire larger weight after normalization. As shown in the next
subsection, the peak position $x_{k_0k_1}^{\max}$ therefore provides a natural
control parameter for adaptive feedback and deterministic convergence.

\subsection{Adaptive Feedback}

To achieve deterministic preparation of the ground state, we introduce
an adaptive feedback strategy based on the measurement record. After
each measurement step, the accumulated outcomes determine the modulation
envelope in the energy eigenbasis, including the peak position
(\ref{eq:xmax}) of the corresponding measurement-induced
energy-filtering profile. This peak position is used as a control
parameter for feedback. If the peak position remains below the threshold
$\epsilon E_{\mathrm{th}}$, the measurement sequence continues without
modification. If the peak position exceeds this threshold, a feedback
operation is applied to redistribute population in the Hamiltonian
eigenbasis and prevent the trajectory from remaining localized in
excited state subspaces.

The iterative evolution of the quantum state is  given by
\begin{equation}
|\psi_{t+1}\rangle
=
\frac{
U_{k_0^{(t+1)}k_1^{(t+1)}} M_n |\psi_t\rangle
}{
\sqrt{\langle \psi_t | M_n^\dagger M_n | \psi_t \rangle}
},
\end{equation}
where $n\in\{0,1\}$ labels the measurement outcome at step $t+1$, with
\begin{equation}
U_{k_0k_1}
=
\begin{cases}
I, & x_{k_0k_1}^{\max}<\epsilon E_{\mathrm{th}},\\[6pt]
U_C, & \mathrm{otherwise}.
\end{cases}
\label{eq:feedbackRule}
\end{equation}
where energy threshold $E_{\mathrm{th}}=(E_0+E_1)/2$ to target the ground state.   To ensure deterministic convergence toward the ground state, the
feedback unitary should have nonzero transition matrix elements between
the relevant energy eigenstates,
\begin{equation}
\left|\langle E_n|U_C|E_m\rangle\right|>0,
\qquad \forall\, n,m ,
\end{equation}
so that population can be redistributed throughout the Hamiltonian
eigenbasis.

\section{Squeezed state preparation}
\label{sec:squeezed}

We now introduce the first of two applications of MITE in bosonic systems.  In this section, we examine the deterministic preparation of squeezed states.

\subsection{Physical system}

We consider the single-mode squeezing Hamiltonian
\begin{equation}
\hat H=\omega \hat a^\dagger \hat a
+\frac{\lambda}{2}\left(\hat a^2+\hat a^{\dagger2}\right),
\label{eq:quadH}
\end{equation}
which is a standard single-mode degenerate parametric-amplifier
Hamiltonian, up to an irrelevant constant energy offset and phase
convention~\cite{Milburn1981,Walls1983,WallsMilburn2008,Colpa1978,Serafini2005}. For $|\lambda|<\omega$,
the Hamiltonian can be diagonalized by a Bogoliubov transformation (see Appendix \ref{app:b} for further details)  
\begin{equation}
\hat b=\cosh r\,\hat a+\sinh r\,\hat a^\dagger ,
\end{equation}
which gives a diagonalized Hamiltonian
\begin{equation}
\hat H=\Omega\, \hat b^\dagger \hat b + E_0,
\label{eq:diagH}
\end{equation}
with
\begin{equation}
\Omega=\sqrt{\omega^2-\lambda^2},
\label{eq:gapOmega}
\end{equation}
where $E_0$ is the ground-state energy. The condition
$|\lambda|<\omega$ ensures that the
spectrum is bounded from below. 

The ground state of Eq.~(\ref{eq:quadH}) is a squeezed vacuum
state characterized by the squeezing parameter $r$, which satisfies
\begin{equation}
\tanh(2r)=\frac{\lambda}{\omega}.
\label{eq:rdef}
\end{equation}
As a result, the quadrature variance is reduced to
\begin{equation}
\langle (\Delta \hat X)^2\rangle=\frac{1}{2}e^{-2r},
\label{eq:Xvar}
\end{equation}
where $\hat X=(\hat a+\hat a^\dagger)/\sqrt{2}$ denotes the
position quadrature operator. The corresponding squeezing strength is commonly expressed in decibels (dB), given as
\begin{equation}
S_{\mathrm{dB}}=-10\log_{10}\!\left(e^{-2r}\right).
\end{equation} 
Such squeezed-vacuum states form a
canonical class of bosonic resource states in continuous-variable
quantum information processing~\cite{Braunstein2005,Weedbrook2012}.
The quasiparticle frequency $\Omega$ sets the characteristic energy
scale of the excitation spectrum and will later be shown to govern
the convergence behavior of the measurement-based imaginary time 
evolution protocol, thereby explaining the degradation of deterministic
preparation near the stability boundary $|\lambda|\rightarrow\omega$.

\subsection{Deterministic preparation with MITE}

We apply MITE to the Hamiltonian (\ref{eq:quadH}) and aim to deterministically prepare squeezed states by evolving the system towards its ground state.  In addition to the Hamiltonian, MITE requires the application of a  unitary  $U_C $ which is applied conditionally upon the measurement outcome (\ref{k0k1state}).  In principle, the feedback operation can be chosen as any non-commuting unitary with respect to the target Hamiltonian $\hat H$,
such that it redistributes the population in the Hamiltonian eigenbasis.
In this work, we choose the feedback operation as
\begin{equation}
U_C=e^{-i\phi\,\hat a^\dagger \hat a},
\label{ucdef}
\end{equation}
which corresponds to a simple single-mode phase rotation. 
This operation helps
prevent measurement trajectories from remaining localized in
excited state subspaces. 

The feedback threshold is chosen from the low-energy
spectrum of the even parity sector as  
\begin{equation}
E_{\mathrm{th}}
=
(E_0
+
E_2
)/2,
\end{equation}
where $E_0$ and $E_2$ denote the
ground and the second excited state. This choice is natural because
the initial vacuum state lies in the even parity sector, and the
Hamiltonian, measurement operators, and feedback operation all preserve
parity symmetry (see Appendix \ref{app:d}).  Thus $ E_1 $ remains unpopulated throughout the evolution, making the second excited state $ E_2 $ a more suitable choice.


\subsection{Convergence with conditional rotation}

\begin{figure}t] \includegraphics[width=\columnwidth]{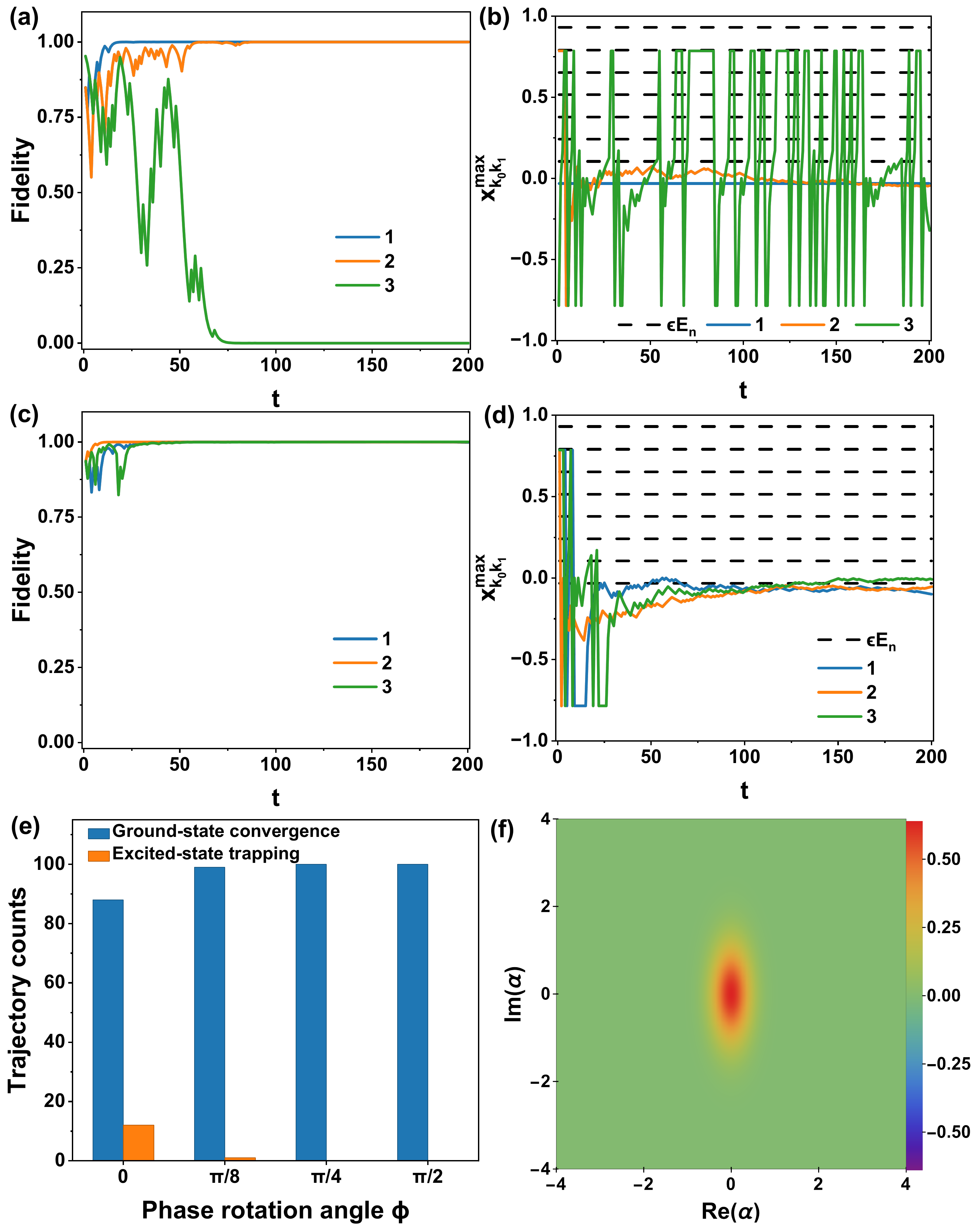} \caption{\label{fig:fig_2} Effect of the phase rotation angle on the MITE evolution of the single-mode squeezing Hamiltonian (\ref{eq:quadH}).  (a),(b)  Fidelity and peak position $x_{k_0k_1}^{\max}$ without phase-rotation feedback, $\phi=0$. (c),(d) Fidelity and peak position $x_{k_0k_1}^{\max}$ with $\phi=\pi/2$. In (b) and (d), the horizontal dashed lines indicate the rescaled
eigenenergies $\epsilon E_n$.(e) Counts of trajectories that converge to the ground state and trajectories that remain trapped in excited states for different rotation angles $\phi$. (f) Wigner function of the prepared state with $\phi=\pi/2$. 
For all calculations we use $\lambda=0.73$, $\epsilon=0.2$, 
$n_{\rm cut}=23$, and $ \omega = 1 $. 
} 
\end{figure}

In this section, we examine how the conditional phase-rotation (\ref{ucdef}) affects
the convergence behavior of the MITE protocol. 

First we consider the case when the conditional unitary $U_C $ is removed entirely (equivalent to setting $\phi=0$ in (\ref{ucdef})).  As shown in Fig.~\ref{fig:fig_2}(a)(b), without the phase-rotation
feedback, some measurement trajectories converge to the ground state,
whereas others remain localized in excited state regions. This behavior is reflected both in the fidelity (Fig.~\ref{fig:fig_2}(a)) and in the peak position $x_{k_0k_1}^{\max}$ (Fig.~\ref{fig:fig_2}(b)). In particular, the trapped
trajectory exhibits a strong fidelity drop, while
$x_{k_0k_1}^{\max}$ fluctuates among the rescaled energy levels indicated
by the dashed lines in Fig.~\ref{fig:fig_2}(b). Thus, repeated weak
measurements alone do not guarantee deterministic convergence. 

With the conditional phase-rotation (\ref{ucdef}) in place, the behavior improves as shown in Fig.~\ref{fig:fig_2}(c)(d). For $\phi=\pi/2$, the representative
trajectories all converge rapidly to high fidelity. When the 
peak exceeds the threshold, the feedback operation redistributes the 
population in the Hamiltonian eigenbasis, preventing the measurement
trajectory from remaining localized in an excited state subspace. Subsequent measurements can then continue the collapse towards the ground state.

The trajectory statistics for different rotation angles are summarized
in Fig.~\ref{fig:fig_2}(e). Without phase-rotation feedback
($\phi=0$), a finite fraction of trajectories remains trapped in
excited states. Increasing $\phi$ suppresses this trapping, and for
$\phi=\pi/4$ and $\phi=\pi/2$ all sampled trajectories converge to the
ground state within numerical resolution. The Wigner function of the prepared state after convergence is shown in
Fig.~\ref{fig:fig_2}(f) for $\phi=\pi/2$. The distribution is centered at
the origin and exhibits the anisotropic Gaussian profile expected for
the squeezed-vacuum ground state of the single-mode squeezing  Hamiltonian,
confirming that the adaptive phase-rotation feedback prepares the
desired bosonic ground state. The corresponding Wigner function of the
ideal ground state of $\hat H$ is visually
indistinguishable from the prepared state on the scale of the plot.

\subsection{Conditional rotation in the energy eigenbasis}

\begin{figure}[b]
\includegraphics[width=0.95\columnwidth]{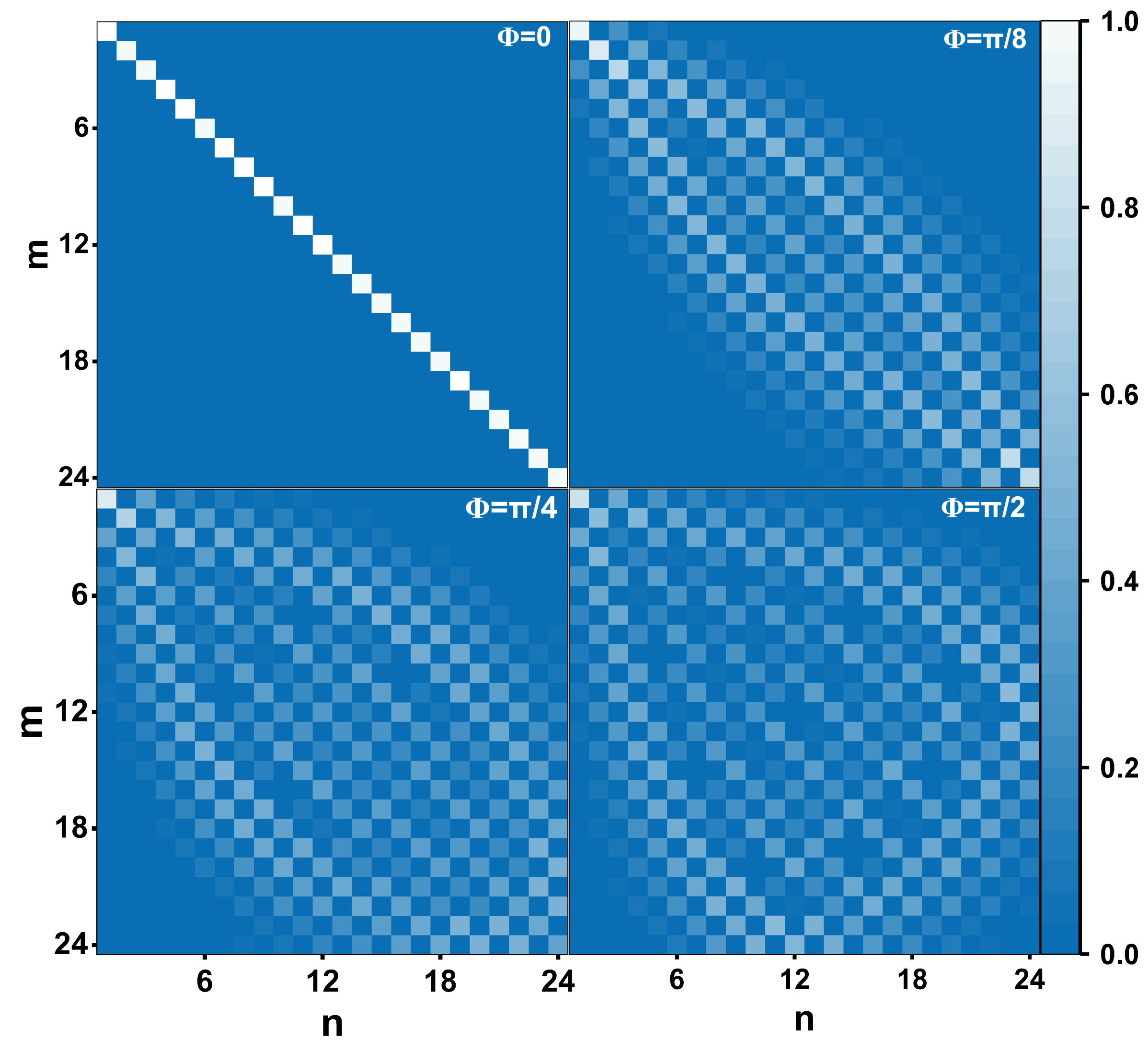}
\caption{\label{fig:fig_3}
The MITE conditional unitary (\ref{ucdef}) in the energy eigenbasis of the single-mode squeezing Hamiltonian (\ref{eq:quadH}). Absolute values of
the matrix elements as given in (\ref{ucmatelem}) are shown for
representative phase-rotation angles $\phi=0,\pi/8,\pi/4$, and
$\pi/2$ as marked. 
}
\end{figure}

To clarify the mechanism behind the conditional rotation dependence, we
represent (\ref{ucdef}) in the energy eigenbasis of the
single-mode squeezing Hamiltonian 
%
\begin{equation}
[U_C]_{mn}
=
\langle E_m|U_C|E_n\rangle ,
\label{ucmatelem}
\end{equation}
where $m$ and $n$ label energy eigenstates ordered by eigenenergy. We plot the absolute value of the matrix elements (\ref{ucmatelem}) for
representative rotation angles in Fig.~\ref{fig:fig_3}.

For $\phi=0$, the conditional operator reduces to the identity, so only
the diagonal elements of $U_C $ are nonzero. In this
case, $ U_C$  leaves each energy eigenstate unchanged and provides
no mechanism for redistributing population away from excited state
regions selected by the measurement record. For finite $\phi$,
off-diagonal elements appear in $U_C $, indicating that
the phase rotation couples different energy eigenstates and reshapes the
state before the subsequent filtering steps. As $\phi$ increases, the
off-diagonal structure becomes broader and more pronounced. The
checkerboard-like pattern reflects the parity structure of the protocol.  Since the single-mode squeezing Hamiltonian and the conditional operation both
preserve parity, the induced mixing remains restricted to states with
the same photon-number parity. Thus, the conditional rotation redistributes
population within each parity sector rather than coupling the even- and
odd-parity subspaces.

This energy-basis structure accounts for the convergence behavior in
Fig.~\ref{fig:fig_2}. When $\phi=0$, the absence of off-diagonal mixing
leaves trajectories susceptible to localization in the excited states.
Increasing $\phi$ broadens the off-diagonal structure within each
parity sector, enabling stronger population redistribution after
conditional rotation is triggered. In the parameter regime considered here,
$\phi=\pi/2$ produces the broadest off-diagonal structure among the
angles tested, consistent with the most robust ground-state convergence.

\subsection{Convergence and the Bogoliubov gap}
\label{sec:boggap}

\begin{figure}[t]
\includegraphics[width=\columnwidth]{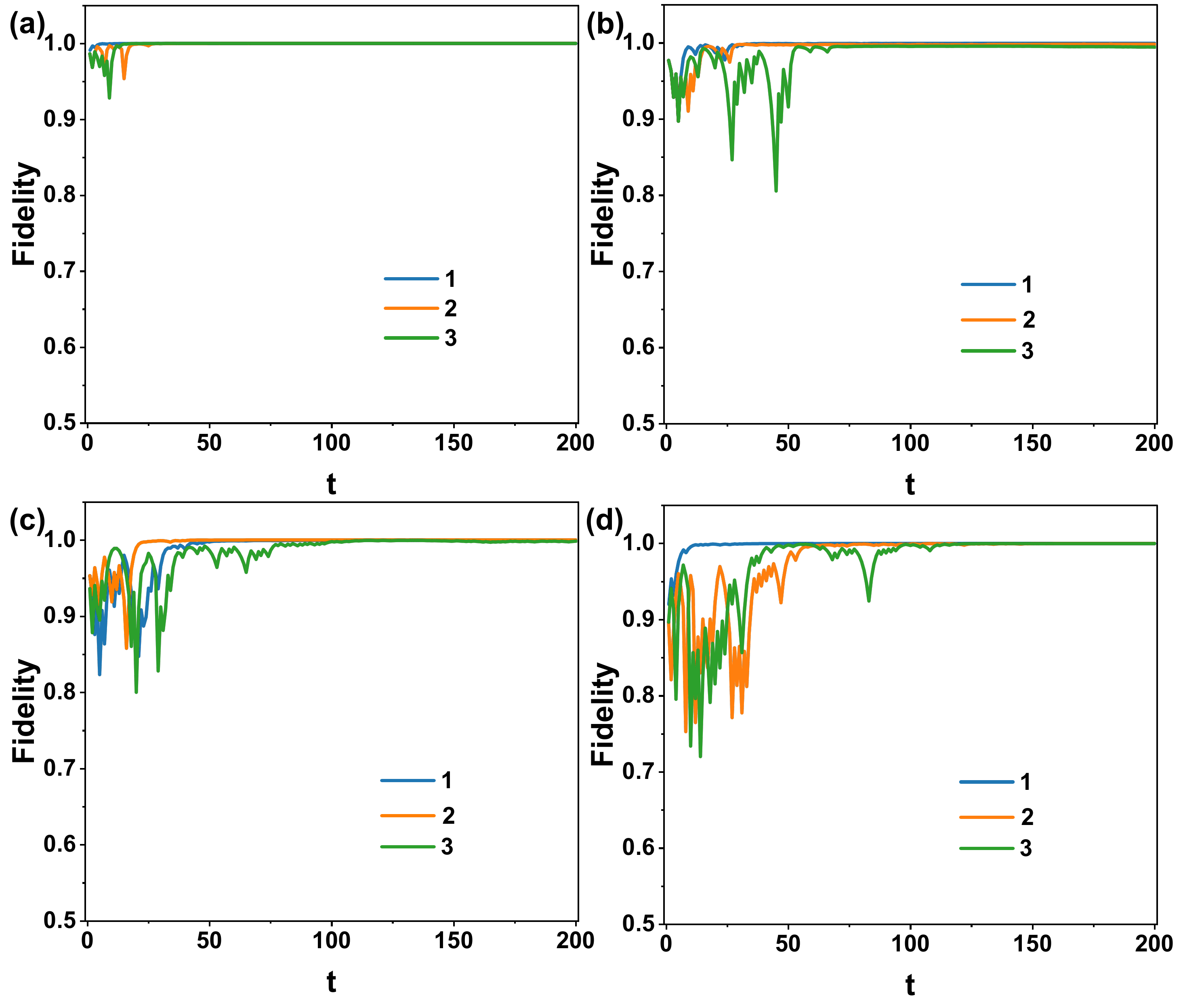}
\caption{\label{fig:fig_4}
Ground state convergence with respect to various squeezing strengths of the single-mode squeezing Hamiltonian (\ref{eq:quadH}).  We use parameters  (a) $\lambda=0.43$ ($S_\mathrm{dB}=2 \text{ dB}$), (b) $\lambda=0.60$ ($S_\mathrm{dB}=3 \text{ dB}$),
(c) $\lambda=0.73$ ($S_\mathrm{dB}=4 \text{ dB} $), and (d) $\lambda=0.82$ ($S_\mathrm{dB}=5 \text{ dB}$). Representative stochastic trajectories of the ground-state fidelity are shown for all cases. For all calculations we use $\phi=\pi/2$, $\epsilon=0.2$, $n_{\rm cut}=23$, and $ \omega = 1 $.
}
\end{figure}

We next examine the effect of increasing the squeezing
strength. Having established that $\phi=\pi/2$ in (\ref{ucdef})  provides the most robust
convergence among the rotation angles tested for the squeezed-vacuum
Hamiltonian, we use this for all calculations henceforth. Our results are shown in Fig.~\ref{fig:fig_4}.   Representative stochastic trajectories
are considered over a range of squeezing strengths from
$2$ to $5\,\mathrm{dB}$. In all cases shown, the MITE protocol
converges to a high-fidelity squeezed vacuum ground state.
For weaker squeezing, the trajectories tend to converge to unit fidelity more quickly,
indicating efficient suppression of excited state components by the
measurement-induced modulation envelope. As the squeezing strength is
increased, the trajectories exhibit more pronounced stochastic
excursions before returning to high fidelity under adaptive
phase-rotation. 

This trend can be understood from the Bogoliubov quasiparticle gap
$\Omega$ defined in Eq.~(\ref{eq:gapOmega}). As the system approaches
the stability boundary $|\lambda|\to\omega$, the gap decreases and the
spectrum density increases. The modulation envelope in the
energy eigenbasis then becomes less able to resolve the ground state
from nearby excitations, making the dynamics more susceptible to
temporary localization in the excited state regions. 
We examine the near-critical regime in more
detail below.

\subsection{Performance limits near the stability boundary}
\label{sec:performlimits}

\begin{figure}[b]
\includegraphics[width=\columnwidth]{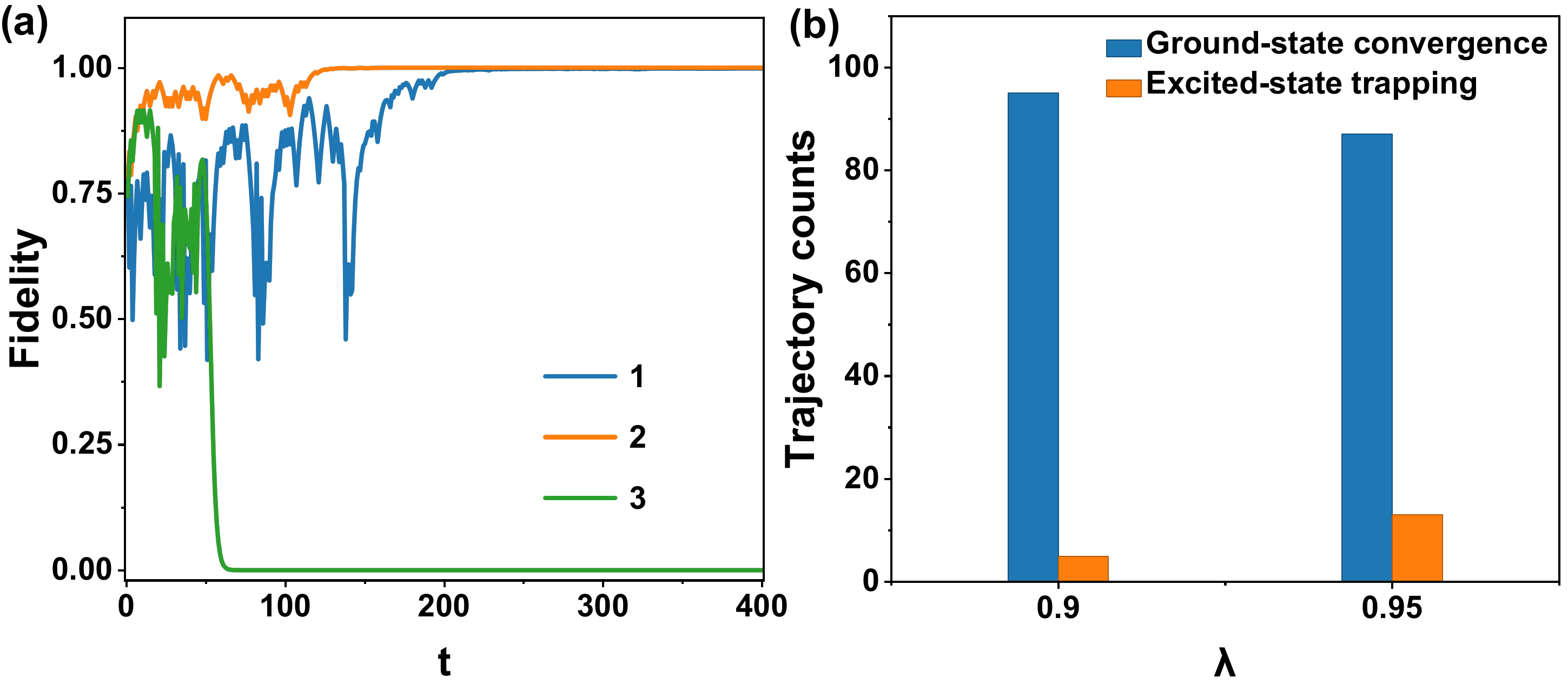}
\caption{\label{fig:fig_5}
Ground state convergence near the stability boundary of the single-mode squeezing Hamiltonian (\ref{eq:quadH}). (a) Fidelity for three representative stochastic quantum trajectories at
$\lambda=0.95$. (b) Counts of trajectories that converge to the ground state and trajectories that remain trapped in excited states for $\lambda=0.9$ and $0.95$.  For all calculations we use $\phi=\pi/2$, $\epsilon=0.2$, $n_{\rm cut}=47$, and $ \omega = 1 $. 
}
\end{figure}

We next examine the protocol close to the stability boundary
$\lambda/\omega \to 1^-$,  where deterministic
convergence is expected to become more difficult, according to the 
discussion of Sec. \ref{sec:boggap}.   Representative fidelity trajectories for $\lambda/\omega =0.95$ are shown in
Fig.~\ref{fig:fig_5}(a). Compared with the moderate squeezing regime, the
convergence is slower and exhibits stronger stochastic fluctuations. Some
trajectories still recover and approach unit fidelity, whereas one
representative trajectory rapidly loses overlap with the target state
and remains localized away from the ground state. The trajectory statistics near the stability boundary are summarized in
Fig.~\ref{fig:fig_5}(b). The protocol still achieves a high convergence
probability, with most sampled trajectories reaching the ground state
for both values of $\lambda$ considered. However, the number of
trajectories trapped in excited states increases as $\lambda$ is raised
from $0.9$ to $0.95$, indicating that the deterministic character of the
protocol weakens as the stability boundary is approached.

\begin{figure}
\includegraphics[width=\columnwidth]{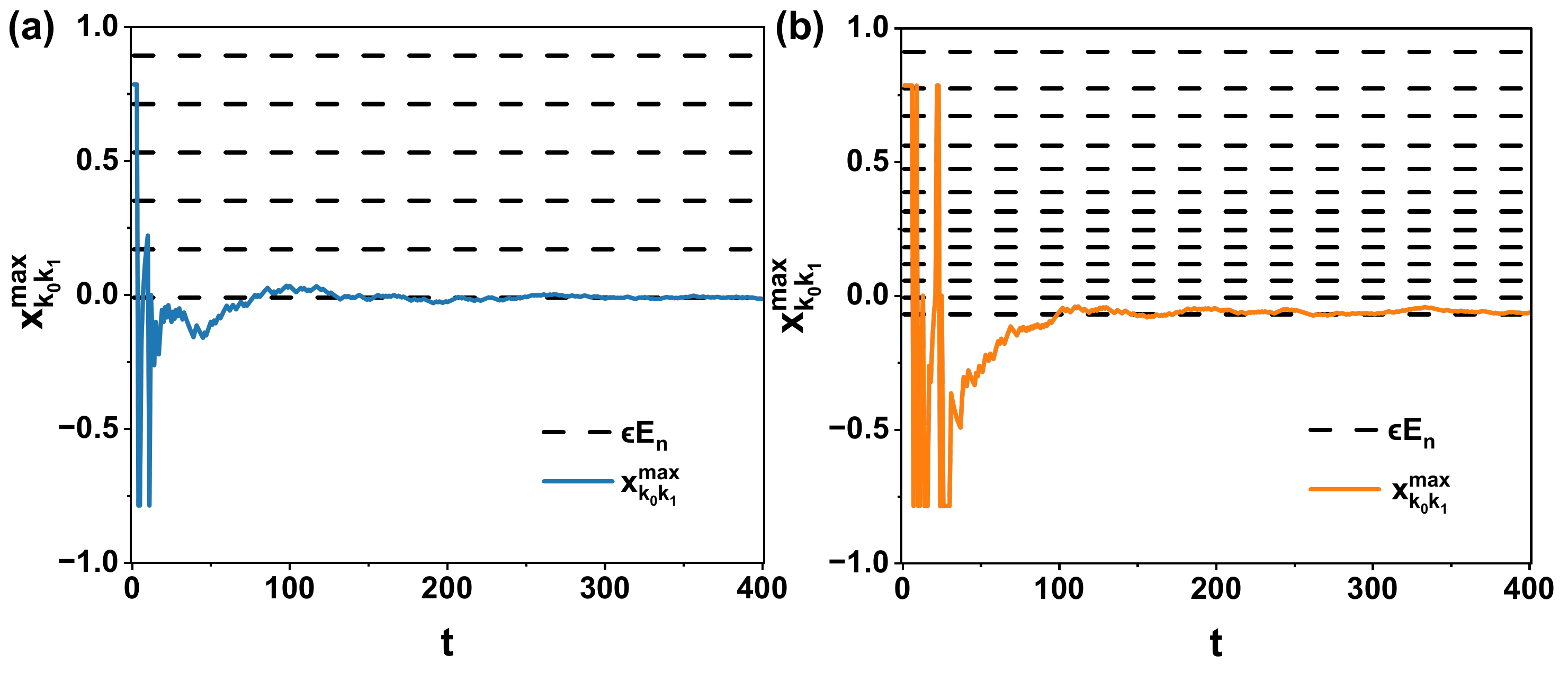}
\caption{\label{fig:fig_6}
Convergence in energy space for the estimated peak position $x_{k_0k_1}^{\max}$ for the single-mode squeezing Hamiltonian (\ref{eq:quadH}).  Parameters chosen are (a) $\lambda=0.43$ and (b)
$\lambda=0.95$.  The horizontal dashed lines indicate the rescaled
eigenenergies $\epsilon E_n$.  For all calculations we use  $\epsilon=0.2$, $\phi=\pi/2$, $n_{\rm cut}=47$, and $ \omega = 1 $.  
}
\end{figure}

The reduced uniformity of convergence near the stability boundary can
be traced to the shrinking Bogoliubov quasiparticle gap $\Omega$, which
increases the density of the spectrum. This mechanism is examined more
directly in Fig.~\ref{fig:fig_6}.  For $\lambda=0.43$, the low-energy levels are well separated on the
scaled energy axis, and the estimated peak energy rapidly returns to the
ground-state region after transient fluctuations. For $\lambda=0.95$,
the low-energy levels are much more closely spaced, and the estimated peak energy exhibits larger and more persistent excursions. 
When the scaled level spacing
$\epsilon (E_{n+1} - E_n) $ is large, the filtering profile can separate the
ground-state region from nearby excitations efficiently. Near the
stability boundary, the shrinking Bogoliubov gap reduces this spacing,
so the filtering profile more easily overlaps with neighboring
excited state regions. Adaptive operations can partially suppress this
drift, but once the low-energy density of states increases such that it cannot be 
cleanly resolved, deterministic convergence gradually degrades.

\subsection{Including bosonic loss}

\begin{figure}[t]
\includegraphics[width=\columnwidth]{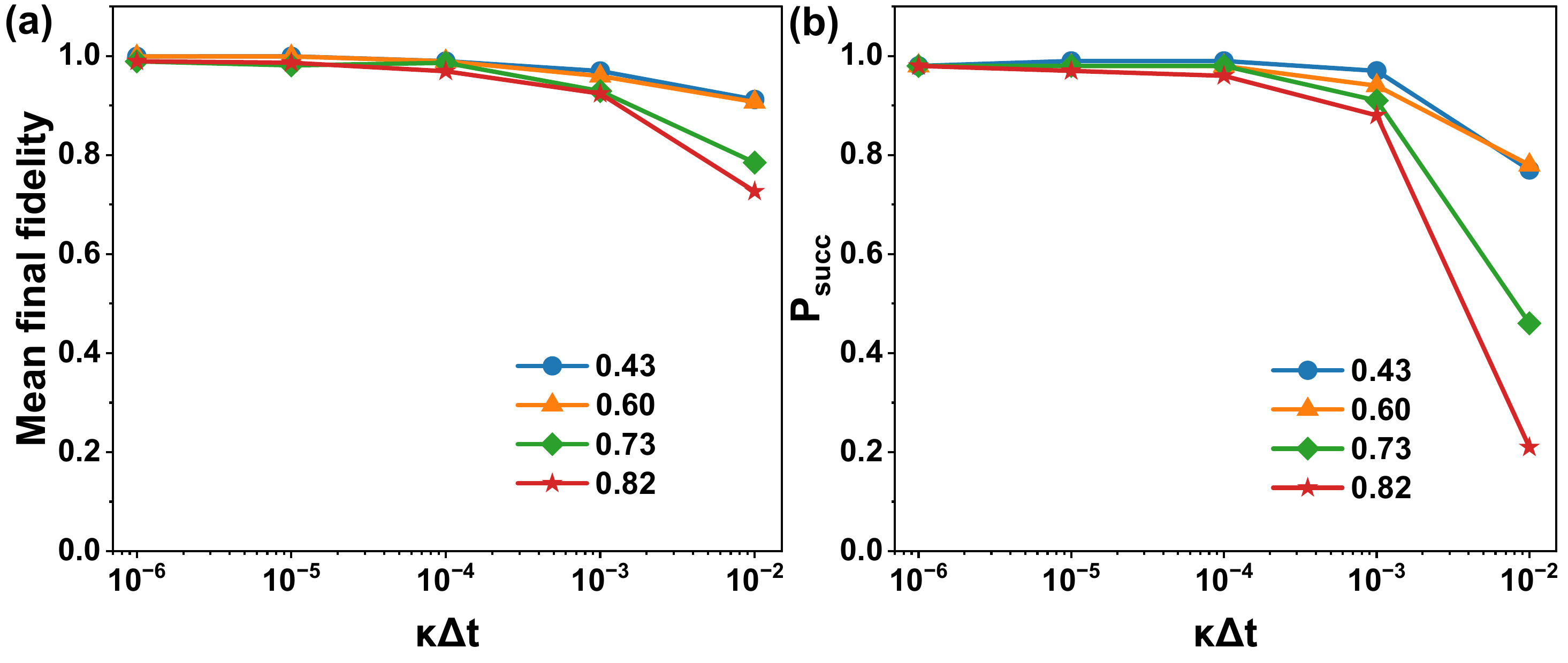}
\caption{\label{fig:fig_7}
Effect of including bosonic loss in the MITE evolution of the single-mode squeezing Hamiltonian (\ref{eq:quadH}).    Bosonic loss is incorporated using quantum jump unraveling of the trajectories, applying (\ref{jumpprob}), (\ref{jumped}), (\ref{notjumped}). 
(a) Mean final fidelity $\overline{F}_{\rm final}$ and (b) success probability $P_{\rm suc}=\Pr(F_{\rm final}>0.99)$ for different squeezing strengths $\lambda$.  We use $ N = 100 $ trajectories in the unraveling.    For all calculations we use $\phi=\pi/2$, $\epsilon=0.2$, $n_{\rm cut}=23$, and $ \omega = 1 $.    
}
\end{figure}

Loss is typically the dominant noise channel in many bosonic systems and is important to consider to connect it to realistic situations. 
In this section, we examine
the robustness of the MITE protocol including bosonic loss. 
To take this into account,  after each
successful weak-measurement update the state is evolved for a short time
step $\Delta t$ under the standard single mode boson loss channel
\cite{WallsMilburn2008,Molmer1996,Fan2014,Schlosshauer2019,Campaioli2024},
\begin{equation}
\frac{d\hat\rho}{dt}
=
\kappa
\left(
2\hat a \hat\rho \hat a^\dagger
-
\hat a^\dagger \hat a \hat\rho
-
\hat\rho \hat a^\dagger \hat a
\right),
\label{eq:photonloss}
\end{equation}
where $\kappa$ is the  bosonic loss rate.

In the wave-function trajectory simulation, this Lindblad channel is
implemented using the corresponding quantum-jump unraveling. For a small
time step $\Delta t$, the jump probability is
\begin{equation}
p_{\rm jump}
=
2\kappa\Delta t
\langle \hat a^\dagger \hat a\rangle .
\label{jumpprob}
\end{equation}
If a jump occurs, the state is updated as
\begin{equation}
|\psi\rangle
\rightarrow
\frac{\hat a|\psi\rangle}{\|\hat a|\psi\rangle\|}.
\label{jumped}
\end{equation}
Otherwise, the no-jump evolution is approximated to first order in
$\Delta t$ as
\begin{equation}
|\psi\rangle
\rightarrow
\frac{
\left(1-\kappa\Delta t\,\hat a^\dagger \hat a\right)|\psi\rangle
}
{
\left\|
\left(1-\kappa\Delta t\,\hat a^\dagger \hat a\right)|\psi\rangle
\right\|
}.
\label{notjumped}
\end{equation}
For each loss strength $\kappa\Delta t$, we simulate $N$  independent
measurement trajectories and record the final fidelity with respect to
the target squeezed ground state. 
A trajectory is counted as successful
when $F_{\rm final}>0.99$, giving the success probability
\begin{equation}
P_{\rm suc}
=
\Pr(F_{\rm final}>0.99).
\end{equation}

The mean final fidelity and the success probability are shown in
Fig.~\ref{fig:fig_7} as functions of the per-step loss strength
$\kappa\Delta t$. Both quantities remain close to their lossless
values for $\kappa\Delta t\lesssim 10^{-4}$, indicating that the
MITE protocol is robust as long as the bosonic
loss is sufficiently small. As the loss strength increases,
both the mean final fidelity and $P_{\rm suc}$ decrease, with a more
visible degradation setting in around $\kappa\Delta t\gtrsim 10^{-3}$.

The loss affects highly squeezed states to a greater extent.  Larger $\lambda$ corresponds to stronger squeezing, which contains a larger photon number population in the target ground
state.  This makes the preparation more sensitive to bosonic loss. In
addition, a boson loss event changes the parity of the
state. Because the target squeezed vacuum lies in the even parity
sector, whereas both the weak measurement operators and the
phase rotation preserve parity, such jumps can drive a
trajectory out of the target sector and hinder its return to the
desired ground state. This effect is most clearly reflected in the
success probability, whose suppression becomes substantially stronger
for larger $\lambda$ at high loss strengths.

\section{Non-Gaussian State Preparation}

In this section, we extend the MITE framework to
non-Gaussian state preparation~\cite{Walschaers2021,Chabaud2023,Su2019,Yadin2018}.

\subsection{Cat state generation}
\label{sec:catstate}

\begin{figure}[b]
\includegraphics[width=\columnwidth]{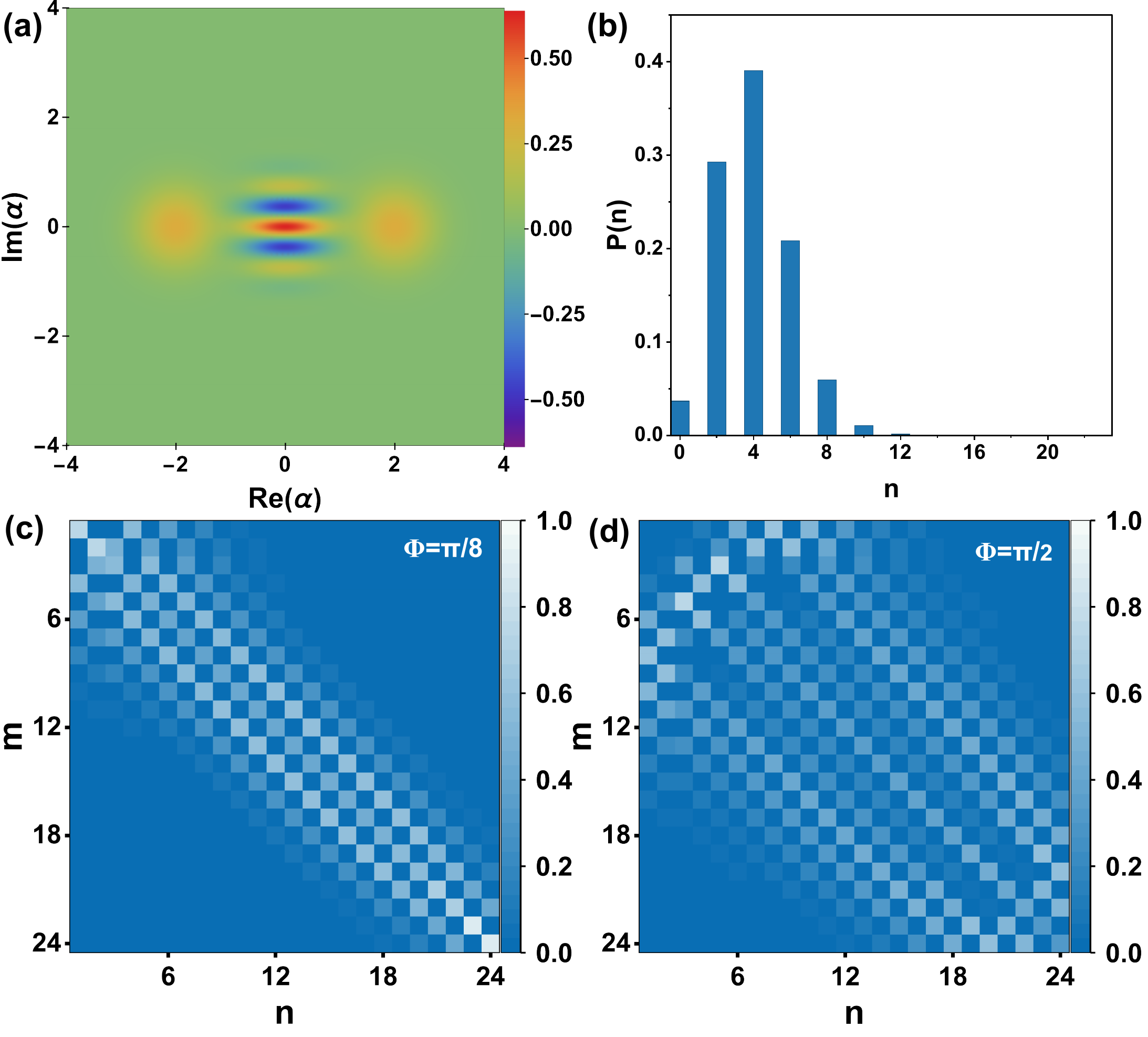}
\caption{\label{fig:fig_8}
Cat state preparation using MITE.  
(a) Wigner function of the prepared state for the cat state parent
Hamiltonian (\ref{eq:Hcat}) with $\alpha=2$.  
(b) Photon-number distribution of the prepared state. 
(c)(d) Absolute values of the matrix elements (\ref{ucmatelem}) 
in the energy
eigenbasis for (c) $\phi=\pi/8$ and (d) $\phi=\pi/2$. 
}
\end{figure}

As a first example of non-Gaussian state generation, we consider Schr\"odinger cat state generation.  We use the cat state parent Hamiltonian \cite{Mirrahimi2014, Su2021, Sychev2017, Bild2023}
\begin{equation}
\hat H_{\mathrm{cat}}
=
(\hat a^{\dagger 2}-\alpha^2)
(\hat a^{2}-\alpha^2) .
\label{eq:Hcat}
\end{equation}
The ground-state manifold is spanned by even- and odd-parity
Schr\"odinger cat states, defined by
\begin{align}
  |C_{\text{even}} \rangle \propto  | \alpha \rangle + | -\alpha \rangle  & = 2 e^{-|\alpha|^2/2} \sum_{n\in \text{even}} \frac{\alpha^n}{\sqrt{n!}} |n \rangle \\
    |C_{\text{odd}} \rangle \propto   | \alpha \rangle - | -\alpha \rangle  & = 2 e^{-|\alpha|^2/2} \sum_{n\in \text{odd}} \frac{\alpha^n}{\sqrt{n!}} |n \rangle .  
\end{align}
Starting from the vacuum state, the protocol
remains confined to the even parity sector because both the
measurement operators and the conditional operation preserve photon number
parity (see Appendix \ref{app:f}). The dynamics therefore selects the even cat
state from the degenerate cat state manifold.

Some representative results for cat state generation with MITE is shown in Fig.~\ref{fig:fig_8}.  
The Wigner function shown in Fig.~\ref{fig:fig_8}(a) exhibits two
coherent state lobes centered near $\pm\alpha$, together with
interference fringes in the central region, which are  the characteristic
phase space signatures of cat states. The
corresponding photon number distribution is dominated by even Fock
components (see Fig.~\ref{fig:fig_8}(b)), consistent with the parity-preserving structure of the
Hamiltonian and the measurement-feedback dynamics.

We then compare the effect of the conditional phase rotation angle in MITE through the energy basis structure of the conditional
operator (see Fig.~\ref{fig:fig_8}(c)(d)).  In contrast to the
squeezed vacuum Hamiltonian, where the larger rotation angle
$\phi=\pi/2$ gives the most robust convergence among the angles tested,
the cat state Hamiltonian is better stabilized by the smaller rotation
angle $\phi=\pi/8$. 
For this parameter regime, $\phi=\pi/8$ yields
deterministic convergence to the even-cat ground state for all sampled
trajectories, whereas $\phi=\pi/2$ leaves a finite probability for
trajectories to remain trapped in excited states. In both cases, the
phase rotation induced mixing remains restricted to the even parity sector,
but the larger rotation angle produces a broader redistribution in the
Hamiltonian eigenbasis.

This difference reflects the distinct structure of the non-Gaussian
target. In the squeezed vacuum case, stronger intra-sector
redistribution is beneficial because it more efficiently suppresses
excited state trapping. The cat state Hamiltonian imposes a different
constraint because the target state is a phase coherent superposition
of multiple even parity Fock components, and an excessively broad
redistribution can compete with the interference structure required for
cat-state formation. The smaller angle $\phi=\pi/8$ therefore provides
a better balance between suppressing trapping and preserving the
coherent structure of the target state.

\subsection{Excited squeezed states}
\label{sec:excited}

\begin{figure}[b]
\includegraphics[width=\columnwidth]{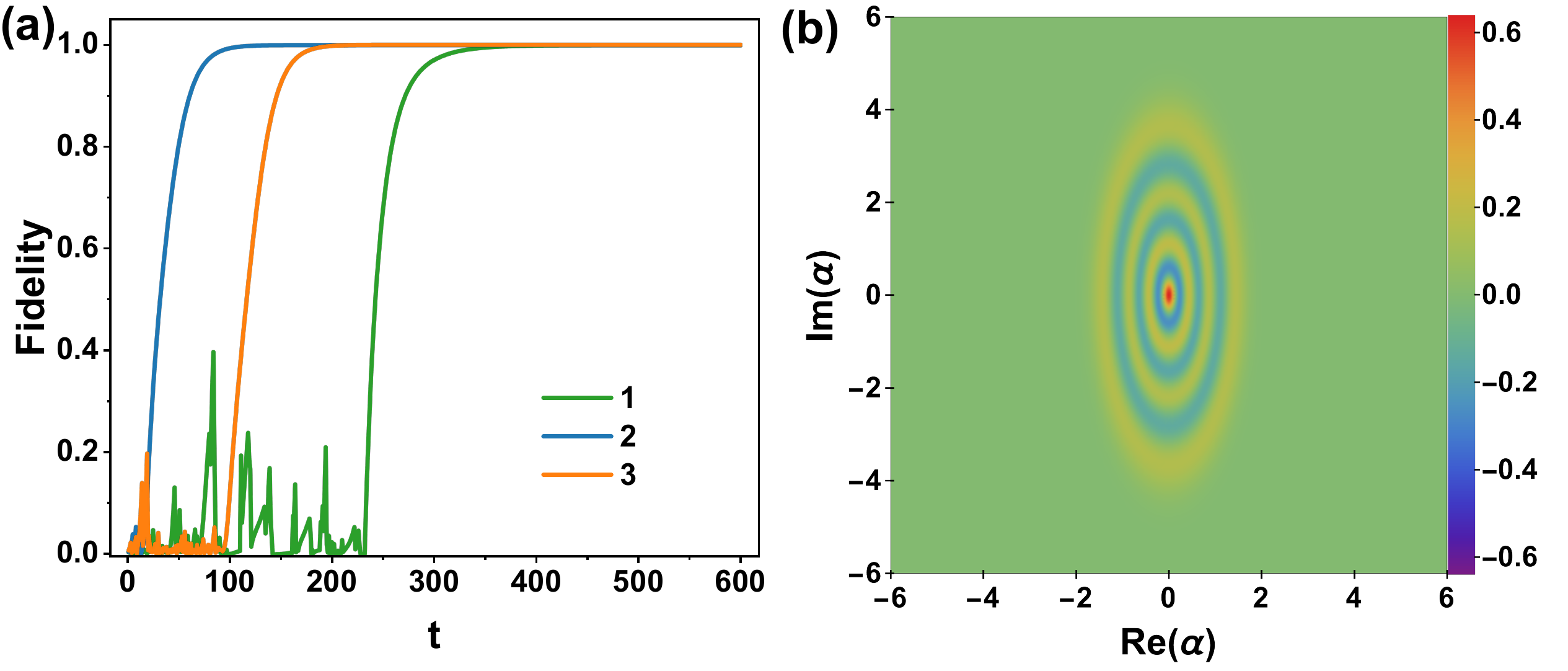}
\caption{\label{fig:fig_9}
Excited squeezed state convergence of the single mode squeezing Hamiltonian (\ref{eq:quadH}).  (a) Fidelity for three representative stochastic quantum trajectories at
$\lambda=0.73$. (b) Wigner function of the prepared state.  For all calculations we use $\phi=\pi/2$, $\epsilon=0.2$, $n_{\rm cut}=47$, and $ \omega = 1 $.  
}
\end{figure}


One of the difficulties of generating non-Gaussian states using the Hamiltonian (\ref{eq:Hcat}) is that it is fourth-order in bosonic operators.  It therefore requires a non-linearity which may be challenging to generate.  In this section, we show another way to generate non-Gaussian states, using only the quadratic single mode squeezing Hamiltonian (\ref{eq:quadH}).  Instead of targeting the ground state, by directly targeting an excited state, non-Gaussian states may be generated.  

We demonstrate this by targeting the sixth excited state  $|E_6\rangle$ of the Hamiltonian (\ref{eq:quadH}).  We set the
feedback threshold to $E_{\mathrm{th}}=(E_4+E_6)/2$, where $E_4$ and
$E_6$ are the two neighboring eigenenergies within the even parity
sector. We modify the ground state feedback rule in
Eq.~(\ref{eq:feedbackRule}) by applying the conditional rotation $U_C$ whenever the peak position defined in Eq.~(\ref{eq:xmax}) satisfies
$x_{k_0k_1}^{\max}<\epsilon E_{\mathrm{th}}$. This reversed feedback
condition prevents the trajectory from remaining localized in
lower energy subspaces and allows the subsequent weak measurements to
favor convergence toward the target excited state (see Appendix \ref{app:g}). 

Representative results for excited state preparation are shown in
Fig.~\ref{fig:fig_9}. Figure~\ref{fig:fig_9}(a) shows the fidelity with
respect to the target state $|E_6\rangle$ for three representative
stochastic trajectories. In contrast to the ground state preparation
considered in Sec. \ref{sec:squeezed}, the initial vacuum state has only a small overlap with
the target excited state, and the initial fidelities are therefore close
to zero. Nevertheless, all three trajectories eventually converge to
near-unit fidelity. The Wigner function of the prepared state is shown in
Fig.~\ref{fig:fig_9}(b). Unlike the single positive Gaussian profile of
the squeezed-vacuum ground state, the excited squeezed state exhibits a
sequence of alternating positive and negative elliptical rings centered
at the origin. The anisotropic deformation of these rings reflects the
underlying quadrature squeezing, while the oscillatory structure and
Wigner negativity arise from the excitation of the target state
$|E_6\rangle$. These features are signatures of a 
non-Gaussian excited eigenstate of the single-mode squeezing
Hamiltonian. 

The above examples show that MITE is not restricted to Gaussian bosonic ground states, but can be extended to non-Gaussian parent Hamiltonians
when the conditional operation is matched to the symmetry and coherence
structure of the target manifold.

\section{SUMMARY AND OUTLOOK}

We have shown that the MITE protocol can be applied to deterministically prepare a variety of
bosonic eigenstates. The central mechanism is
a measurement-feedback loop in which repeated weak measurements
generate a measurement-based modulation envelope in the
energy eigenbasis of the target Hamiltonian. As the number of
measurements increases, this envelope becomes increasingly selective in
energy space, while the adaptive phase rotation redirects
trajectories if the energy does not lie in the specified region.  In this way, stochastic measurement outcomes are converted into deterministic 
convergence toward the target ground state.

For the single mode squeezing Hamiltonian (Sec. \ref{sec:performlimits}), the numerical results show that the
effectiveness of this mechanism is ultimately limited by spectral
resolution. When the Bogoliubov quasiparticle gap becomes small, the density of states increases and the measurement induced envelope
can no longer resolve the ground state from nearby excitations as
efficiently. Deterministic convergence then gradually degrades. We also
showed that the prepared squeezed states remain robust under moderate
boson loss during preparation, with stronger
degradation for larger squeezing arising from the higher excitation number
content of the target states. Beyond Gaussian targets, the same
measurement-feedback framework extends to generation of non-Gaussian states.  In Sec. \ref{sec:catstate} we showed that even cat states could be reliably generated using a fourth order parent Hamiltonian.  The results highlighted that the
conditional operation must be matched to the symmetry and coherence structure of the target state.  Another approach to generating non-Gaussian states is to target excited states (see Sec. \ref{sec:excited}).  This has the advantage that the parent Hamiltonian only has quadratic, rather than fourth order in bosonic operators.

A natural next step is to
extend the same framework to multimode bosonic systems. In that setting,
the main challenge will be to design feedback operations that steer the
measurement-induced envelope within a much larger low energy manifold
while preserving the relevant symmetries and correlations of the target
state. The main obstacle that we anticipate in such multimode systems is the slower convergence time towards the target state.  One advantage of the current scheme is that it is a non-unitary procedure, such that the runtime does not necessarily need to occur within the coherent time of the quantum system.  Coherence is generated by the projection into the Hamiltonian eigenbasis, rather than the coherence generated by the unitary operations.  
Deterministic
preparation of multipartite continuous-variable resource states are potentially useful for applications such as multiparameter quantum metrology, continuous variable quantum communication, bosonic quantum error correction, and quantum networks, to name a few examples.  Investigations in this direction may further broaden the scope of bosonic quantum state
engineering.

\begin{acknowledgments}
This work is supported by the SMEC Scientific Research Innovation Project (2023ZKZD55); the National Natural Science
Foundation of China (92576102); the Science and Technology
Commission of Shanghai Municipality (22ZR1444600); the
NYU Shanghai Boost Fund; the China Foreign Experts Program (G2021013002L); the NYU-ECNU Institute of Physics at
NYU Shanghai; the NYU Shanghai Major-Grants Seed Fund; and Y.M. is supported by the Natural Science
Foundation of Zhejiang Province under Grant No.
LQN26A050002.
\end{acknowledgments}


\appendix

\section{Weak-Measurement Operators and Gaussian Energy Filtering}

\label{app:a}

In the energy eigenbasis of $\hat H$, the measurement
operators in Eqs.~(\ref{eq:M0}) and (\ref{eq:M1}) take the form
\begin{equation}
M_0=
\frac{1}{\sqrt{2}}
\sum_n
\left(\cos(\epsilon E_n)-\sin(\epsilon E_n)\right)
|E_n\rangle\langle E_n|,
\end{equation}
\begin{equation}
M_1=
\frac{1}{\sqrt{2}}
\sum_n
\left(\cos(\epsilon E_n)+\sin(\epsilon E_n)\right)
|E_n\rangle\langle E_n|.
\end{equation}
They satisfy the completeness relation
\begin{equation}
M_0^\dagger M_0+M_1^\dagger M_1=I .
\end{equation}
The imaginary time expressions $ \approx e^{\pm \epsilon \hat{H}}/\sqrt{2} $ are accurate for
$\|\epsilon \hat H\|\ll 1$. The parameter $\epsilon$ is chosen such that the relevant
spectrum lies within
$-\pi/4\le \epsilon E_n\le \pi/4$, where the Hamiltonian was shifted by a suitable energy offset. This range is chosen such that the filtering function $ A $ is non-periodic, such that only one energy state is targeted. 

For a particular measurement sequence consisting of $k_0$ applications
of $M_0$ and $k_1$ applications of $M_1$, the corresponding amplitude
function acting on an energy eigenstate $|E\rangle$ is
\begin{equation}
A_{k_0k_1}(\epsilon E)
=
\frac{\left[\cos(\epsilon E)-\sin(\epsilon E)\right]^{k_0}
\left[\cos(\epsilon E)+\sin(\epsilon E)\right]^{k_1}}{\sqrt{2^{k_0+k_1}}}
.
\end{equation}
Setting $x=\epsilon E$, this can be written as
\begin{equation}
A_{k_0k_1}(x)
=
\cos^{k_0}\left(x+\frac{\pi}{4}\right)
\sin^{k_1}\left(x+\frac{\pi}{4}\right).
\end{equation}
This amplitude function defines the modulation envelope in the energy
eigenbasis.

For a large total number of applications,
$N=k_0+k_1$, expanding the logarithm of
the modulation envelope around its maximum gives a Gaussian-like form,
\begin{equation}
A_{k_0k_1}(x)
\propto
\exp\!\left[
-\frac{\left(x-x_{k_0k_1}^{\max}\right)^2}{2\sigma^2}
\right],
\end{equation}
where the peak position is given by Eq. (\ref{eq:xmax}), and the approximate width is Eq. (\ref{eq:sigma}).

\section{The Single-Mode Squeezing Hamiltonian}

\label{app:b}

Here we provide the derivation for diagonalizing the single-mode squeezing
Hamiltonian in Eq.~(\ref{eq:quadH}). We introduce the Bogoliubov
transformation
\begin{equation}
\hat b=\cosh r\,\hat a+\sinh r\,\hat a^\dagger ,
\end{equation}
which obeys  bosonic commutation relations
\begin{equation}
[\hat b,\hat b^\dagger]
=
\cosh^2 r-\sinh^2 r
=
1 .
\end{equation}
Requiring the off-diagonal terms of the transformed Hamiltonian to
vanish gives
\begin{equation}
\tanh(2r)=\frac{\lambda}{\omega}.
\end{equation}
Equivalently,
\begin{equation}
r=\frac{1}{4}\ln\left(\frac{\omega+\lambda}{\omega-\lambda}\right).
\end{equation}
The Hamiltonian is then diagonalized as
\begin{equation}
\hat H=\Omega\,\hat b^\dagger\hat b+E_0,
\end{equation}
where
\begin{equation}
\Omega=\sqrt{\omega^2-\lambda^2}.
\end{equation}
The stability condition $|\lambda|<\omega$ ensures that $\Omega$ is
real and that the spectrum is bounded from below.



We next discuss the characteristics of this solution for various parameters.  It is illustrative to transform to quadrature variables, defined as
\begin{equation}
\hat x
=
\frac{\hat a+\hat a^\dagger}{\sqrt{2}},
\qquad
\hat p
=
\frac{\hat a-\hat a^\dagger}{i\sqrt{2}},
\end{equation}
or equivalently
\begin{equation}
\hat a
=
\frac{\hat x+i\hat p}{\sqrt{2}},
\qquad
\hat a^\dagger
=
\frac{\hat x-i\hat p}{\sqrt{2}}.
\end{equation}
The quadratures satisfy the canonical commutation relation
\begin{equation}
[\hat x,\hat p]=i .
\end{equation}

First, the number operator can be written as
\begin{align}
\hat a^\dagger \hat a
&=
\frac{1}{2}
(\hat x-i\hat p)(\hat x+i\hat p)
\nonumber\\
&=
\frac{1}{2}
\left(
\hat x^2+\hat p^2-1
\right).
\label{eq:app_number_quad}
\end{align}
Second, the anomalous quadratic term becomes
\begin{align}
\hat a^2
&=
\frac{1}{2}
(\hat x+i\hat p)^2
\nonumber\\
&=
\frac{1}{2}
\left(
\hat x^2
+i\hat x\hat p
+i\hat p\hat x
-\hat p^2
\right),
\\
\hat a^{\dagger 2}
&=
\frac{1}{2}
(\hat x-i\hat p)^2
\nonumber\\
&=
\frac{1}{2}
\left(
\hat x^2
-i\hat x\hat p
-i\hat p\hat x
-\hat p^2
\right).
\end{align}
Adding these two expressions gives
\begin{equation}
\hat a^2+\hat a^{\dagger 2}
=
\hat x^2-\hat p^2 .
\label{eq:app_anomalous_quad}
\end{equation}
Substituting Eqs.~\eqref{eq:app_number_quad} and
\eqref{eq:app_anomalous_quad} into Eq.~\eqref{eq:quadH}, we obtain 
\begin{align}
\hat H
&=
\frac{\omega}{2}
\left(
\hat x^2+\hat p^2-1
\right)
+
\frac{\lambda}{2}
\left(
\hat x^2-\hat p^2
\right)
\nonumber\\
&=
\frac{\omega+\lambda}{2}\hat x^2
+
\frac{\omega-\lambda}{2}\hat p^2
-
\frac{\omega}{2}.
\label{eq:app_quadH_xp}
\end{align}

Equation~\eqref{eq:app_quadH_xp} shows that the Hamiltonian describes a
stable harmonic oscillator only when both quadratic coefficients are
positive:
\begin{equation}
\omega+\lambda>0,
\qquad
\omega-\lambda>0.
\end{equation}
For $\omega>0$, this gives
\begin{equation}
|\lambda|<\omega .
\label{eq:app_stability_condition}
\end{equation}
In this regime, the Hamiltonian has a discrete spectrum and a
normalizable ground state.

For $|\lambda|>\omega$, one of the two quadratic coefficients becomes
negative. If $\lambda>\omega$, the coefficient of $\hat p^2$ is negative,
and the energy expectation value can be made arbitrarily negative by
considering states with increasingly large $\langle \hat p^2\rangle$.
Similarly, if $\lambda<-\omega$, the coefficient of $\hat x^2$ is
negative, and the energy can be made arbitrarily negative by increasing
$\langle \hat x^2\rangle$. Therefore, for $|\lambda|>\omega$, the
 Hamiltonian is unbounded from below.

At the boundary $|\lambda|=\omega$, one of the quadratic coefficients
vanishes and the Bogoliubov excitation gap closes. The normal-mode
description then becomes singular, corresponding to an infinitely
squeezed, non-normalizable limiting state rather than a stable oscillator
with a discrete spectrum.

\section{Parity conservation in the single-mode squeezing Hamiltonian}

\label{app:d}

The feedback threshold in Eq.~(\ref{eq:feedbackRule}) is chosen from
the even parity sector because the ideal protocol preserves parity
throughout the evolution. Define the parity operator
\begin{equation}
\Pi=e^{i\pi \hat a^\dagger \hat a}.
\end{equation}
For the single-mode squeezing Hamiltonian in Eq.~(\ref{eq:quadH}), both the
number-conserving term and the pair-creation and pair-annihilation
terms preserve photon-number parity, and therefore
\begin{equation}
[\hat H,\Pi]=0.
\end{equation}
Since the measurement operators $M_0$ and $M_1$ are functions of
$\hat H$, they also commute with $\Pi$. The feedback operation (\ref{ucdef}) is diagonal in the Fock basis and also satisfies
\begin{equation}
[U_C,\Pi]=0.
\end{equation}
Thus, for an initial vacuum state, the ideal evolution remains
confined to the even parity sector. The relevant threshold is therefore set by the lowest excitation
scale within this symmetry sector.




\section{Parity conservation in the cat state Hamiltonian}
\label{app:f}

The cat-state Hamiltonian in Eq.~(\ref{eq:Hcat}) preserves photon-number
parity,
\begin{equation}
[\hat H_{\mathrm{cat}},\Pi]=0,
\qquad
\Pi=e^{i\pi \hat a^\dagger \hat a}.
\end{equation}
Since the measurement operators are functions of
$\hat H_{\mathrm{cat}}$, they also commute with $\Pi$. The feedback
rotation (\ref{ucdef})
is diagonal in the Fock basis and likewise preserves parity.

Starting from the vacuum state, the evolution therefore remains confined
to the even parity sector for every measurement trajectory. This
symmetry constraint explains why the protocol selects the even cat state
from the degenerate even--odd cat manifold and why odd Fock components
remain absent in the final photon-number distribution.

\section{Excited state preparation}
\label{app:g}

\begin{figure}[t]
\includegraphics[width=0.5\columnwidth]{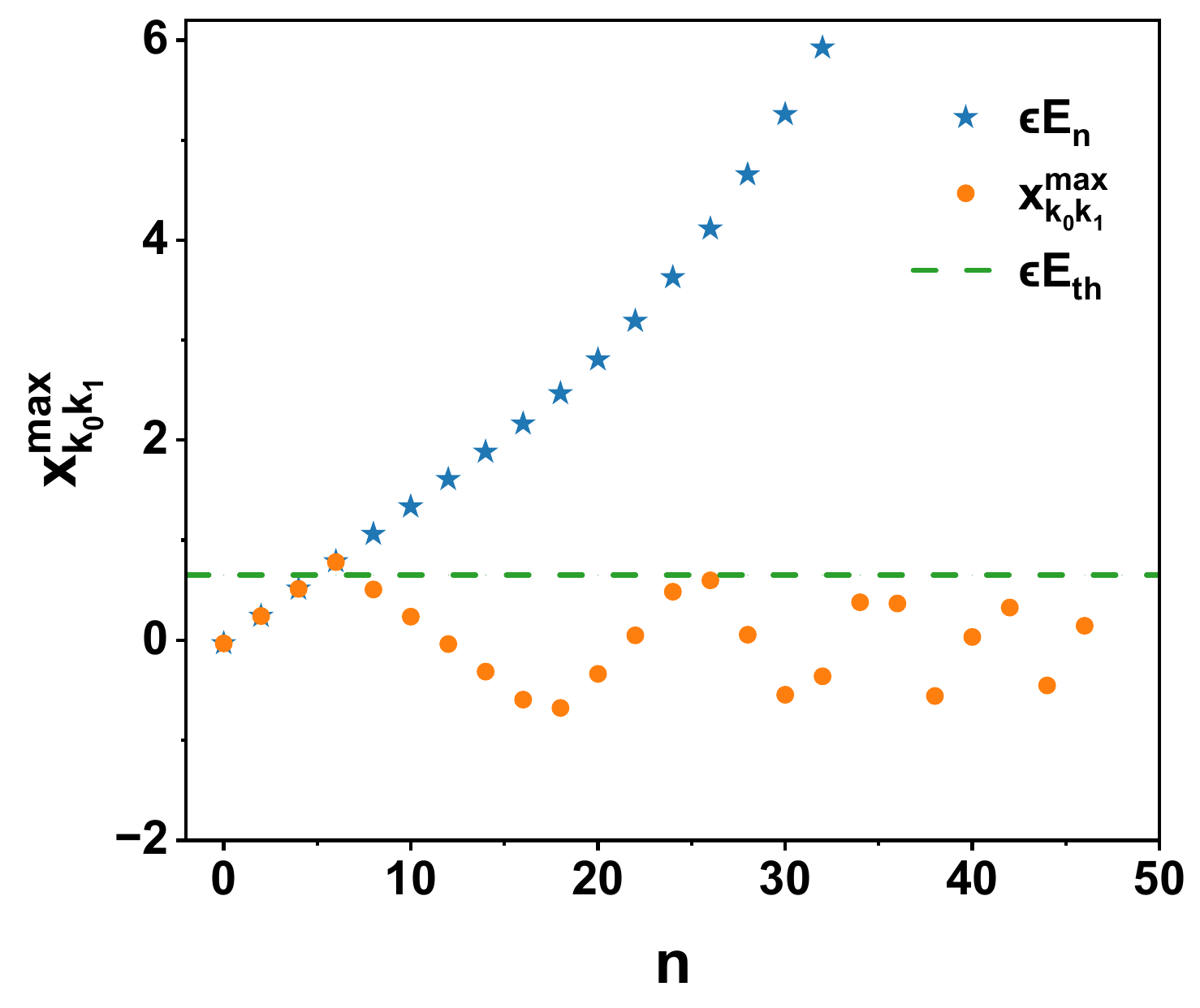}
\caption{\label{fig:fig_13}
Rescaled even-parity eigenenergies $\epsilon E_n$ and estimated peak position $x_{k_0k_1}^{\max}$ for the single-mode squeezing Hamiltonian (\ref{eq:quadH}). For all calculations we use  $\epsilon=0.2$, $\lambda=0.73$, $n_{\rm cut}=47$, and $ \omega = 1 $.  
}
\end{figure}

Combining Eqs.~(\ref{eq:Ak0k1}) and (\ref{eq:xmax}), the peak position associated with the
eigenenergy $E_n$ is given by
\begin{equation}
x_{k_0k_1}^{\max}
=
\frac{1}{2}\arcsin\!\left[\sin(2\epsilon E_n)\right].
\label{eq:folded_peak}
\end{equation}
Since the principal value of $\arcsin$ lies within
$[-\pi/2,\pi/2]$, once $\epsilon E_n$ exceeds $\pi/4$, the inferred
peak position is folded back into the interval
$[-\pi/4,\pi/4]$. As shown in Fig.~\ref{fig:fig_13},  within the
truncated even parity spectrum, the folded peak associated with $E_6$
is the only one satisfying
$x_{k_0k_1}^{\max}>\epsilon E_{\mathrm{th}}$. All other folded peak
positions lie below the threshold and therefore continue to apply 
the conditional rotation $U_{\mathrm{C}}$.

\begin{figure}[b]
\includegraphics[width=\columnwidth]{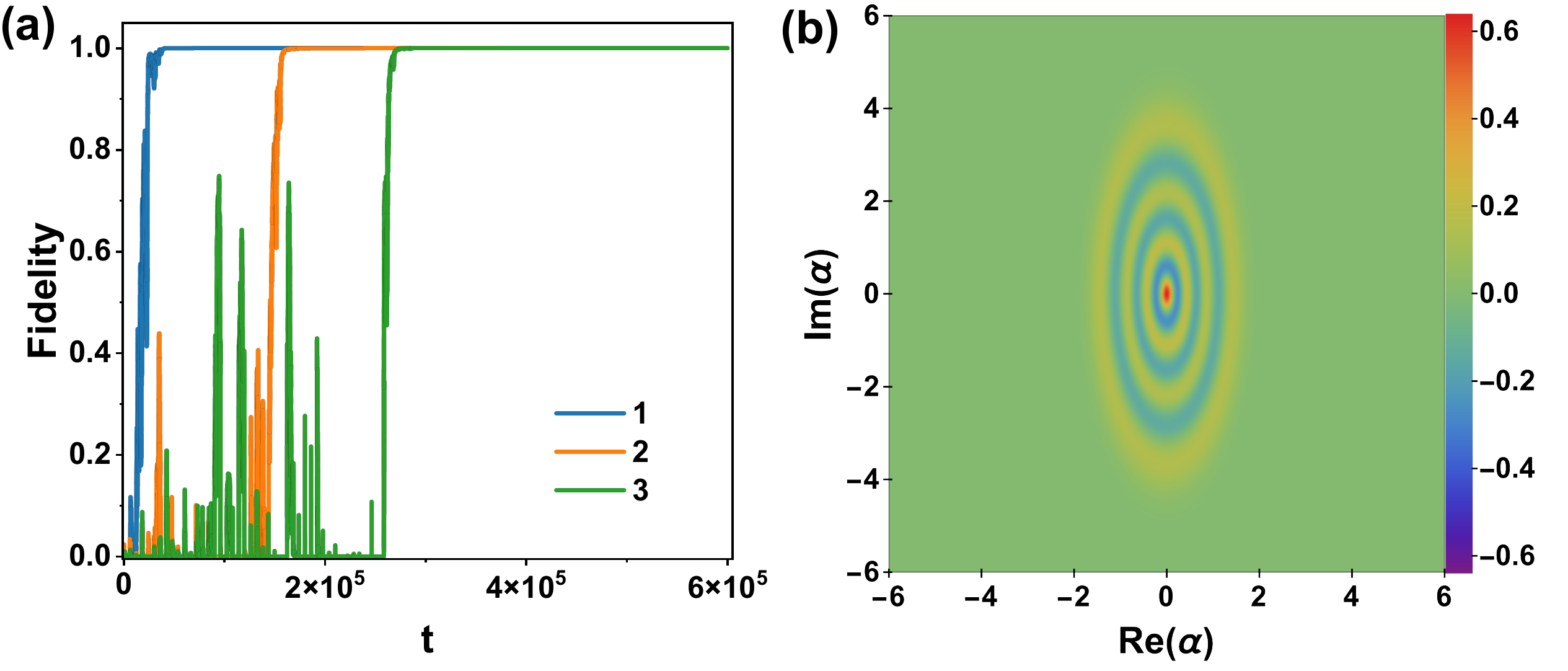}
\caption{\label{fig:fig_12}
Excited squeezed state convergence (a) Fidelity for three representative stochastic quantum trajectories at
$\lambda=0.73$. (b) Wigner function of the prepared state.  For all calculations we use $\phi=\pi/2$, $\epsilon=0.01$, $t_{\min}=5000$, and $n_{\rm cut}=47$. 
}
\end{figure}

Alternatively, we choose a sufficiently small $\epsilon$ such that the
target and its neighboring rescaled eigenenergies lie within the
principal interval $[-\pi/4,\pi/4]$. We then set the two feedback
thresholds to $E_{\mathrm{th},1}=(E_4+E_6)/2$ and
$E_{\mathrm{th},2}=(E_6+E_8)/2$. After a minimum number of measurement
steps ($t_{\min}$) has been accumulated, the conditional rotation
$U_{\mathrm{C}}$ is applied whenever the estimated peak lies outside
the target window,
$\epsilon E_{\mathrm{th},1}<x_{k_0k_1}^{\max}
<\epsilon E_{\mathrm{th},2}$, while no feedback is applied inside this
window. This strategy also prepares
the target state $|E_6\rangle$ shown in Fig.~\ref{fig:fig_12}
with a Wigner function indistinguishable from that shown in Fig.~\ref{fig:fig_9}(b). However, this method requires a longer measurement sequence.

\nocite{*}

\bibliography{apssamp}

\end{document}